\documentclass[12pt]{iopart}

\usepackage{graphicx}% Include figure files
\usepackage{xfrac}
\usepackage{epsfig}
\usepackage{iopams}
\usepackage{color}
\usepackage[amssymb]{SIunits}

\begin{document}

\title[Nonlinear resonator chain]{Networks of nonlinear superconducting transmission line resonators}

\author{M Leib$^1$, F Deppe$^{1,2}$, A Marx$^2$, R Gross$^{1,2}$ and M J Hartmann$^1$}
\address{1 Technische Universit{\"a}t M{\"u}nchen, Physik Department, James-Franck-Str., D-85748 Garching, Germany}
\address{2 Walther-Mei{\ss}ner-Institut, Bayerische Akademie der Wissenschaften, Walther-Mei{\ss}ner-Strasse 8, D-85748 Garching, Germany}
\eads{\mailto{MartinLeib@circuitqed.net}, \mailto{mh@tum.de}}

\date{\today}

\begin{abstract}
We investigate a network of coupled superconducting transmission line resonators, each of them made nonlinear with a capacitively shunted Josephson junction coupling to the odd flux modes of the resonator. The resulting eigenmode spectrum shows anticrossings between the plasma mode of the shunted junction and the odd resonator modes. Notably, we find that the combined device can inherit the complete nonlinearity of the junction, allowing for a description as a harmonic oscillator with a Kerr nonlinearity. Using a dc~SQUID instead of a single junction, the nonlinearity can be tuned between $\unit{10}{\kilo\hertz}$ and $\unit{4}{\mega\hertz}$ while maintaining resonance frequencies of a few gigahertz for realistic device parameters. An array of such nonlinear resonators can be considered a scalable superconducting quantum simulator for a Bose-Hubbard Hamiltonian. The device would be capable of accessing the strongly correlated regime and be particularly well suited for investigating quantum many-body dynamics of interacting particles under the influence of drive and dissipation.
\end{abstract}


%\submitto{\NJP}
\maketitle 

\tableofcontents
\newpage

\section{Introduction}

In present-day research the physics of quantum many-body systems has gained a lot of attention. Many-body Hamiltonians are investigated as minimal models for collective physical phenomena ranging from quantum phase transitions
to quantum transport and nonequilibrium dynamics. Moreover, technological applications are being investigated for systems that can be described by interacting many-body Hamiltonians, such as topologically protected quantum states, novel schemes for quantum error correction or one way quantum computing.

There are only few situations where analytical solutions for interacting many-body Hamiltonians can be found. Normally one would use advanced numerical techniques to investigate theoretically the properties of the many-body Hamiltonian, but numerical methods are limited by the exponential growth of the Hilbert space of many-body systems as one increases the number of particles.
As an alternative approach to circumvent this dilemma, quantum simulators are now being explored intensively.
These are well controllable quantum systems that may emulate the physics of other systems which are less amenable to experimental investigation and are therefore expected to give rise to interesting and yet scarcely explored physics.
Present day quantum simulators are therefore often strongly magnified versions of the respective system one is interested in, where for example lattice constants greatly exceed interatomic distances in a crystal. This can either allow to measure spatial correlations or perform manipulations for specific tasks such as quantum computing. Quantum many-body Hamiltonians can for example be simulated with cold atoms trapped by laser fields in various shapes and dimensions \cite{review_Bloch_2008}, in ion traps \cite{Friedenauer:2008fk,Kim2010} or arrays of cavity quantum electrodynamics (QED) systems \cite{Hartmann:2006kx,hartmann-2008-2,PhysRevA.76.031805,Greentree:2006uq}.

In the latter, instead of massive particles, joint excitations of the cavity field mode and a nonlinear quantum system, so-called polaritons, interact with each other \cite{Hartmann:2006kx,hartmann-2008-2,PhysRevA.76.031805,Greentree:2006uq,1367-2630-10-3-033011,PhysRevA.81.021806,PhysRevA.82.033813,Hartmann2010}.  In this context, the nonlinearity \cite{brandao,PhysRevLett.99.103601} can be regarded as an interaction between polaritons in the same cavity.
Multiple cavities can be coupled to form a network, such that photons can hop between different cavities and consequently the polaritons are capable of propagating through the network of cavities.  Since polaritons are joint excitations of two degrees of freedom, the cavity and the nonlinear quantum system, typically two species of polaritons emerge which are separated in energy by the coupling between cavity and the nonlinear quantum system. For the existence and addressability of the polaritons it is therefore necessary to operate in the strong coupling regime, which can be reached by increasing the coupling between the nonlinear system and the cavity until it exceeds the relevant decay rates. This condition is nowadays extremely well met in the circuit analog of cavity QED, superconducting circuit QED.
In circuit QED microwave photons confined by quasi one-dimensional transmission line resonators are coupled to Josephson junction based nonlinear quantum systems (qubits) \cite{PhysRevA.69.062320,Wallraff:2004rz,Deppe:2008kx}. The very small mode volume of the one-dimensional transmission line and the very large dipole moment of the Josephson qubit result in exceptionally large coupling rates \cite{Niemczyk} while the superconducting gap ensures low dissipation. 

In this work, we propose an even simpler approach, which requires less control circuitry and therefore facilitates coupling multiple resonators to form scalable networks. Instead of employing joint excitations of the resonator and the qubit, a transmission line resonator can directly be rendered nonlinear by intersecting it with a Josephson junction \cite{Mallet:2009fk}. The differences between the eigenfrequencies of the combined resonator - Josephson junction system are here comparable to the fundamental mode frequency, which enables a clear separation between the eigenmodes.
As a consequence, neither the nonlinearity of the modes nor a photon tunneling between adjacent resonators of a network lead to appreciable coupling between different eigenmodes. This leads to a system where the excitations of one eigenmode are dynamically well separated from those of other eigenmodes and can serve as a reliable, high precision quantum simulator.
 
 Transmission lines that are intersected by Josephson junctions \cite{Bourassa} are currently being used for measurement \cite{Mallet:2009fk} or amplification \cite{Lehnert,Majedi} purposes in the strong driving limit. Here we show that the modes of such a device can achieve a significant amount of nonlinearity on the single photon level for realistic circuit parameters. If several of such transmission line resonators are connected to form a network, the dynamics of the photons propagating through this network is thus described by a Bose-Hubbard Hamiltonian. In this way our approach leads to quantum simulations of the physics of massive, interacting particles by employing massless photons. With present day circuit QED technology it has been demonstrated that hundreds of resonators can be fitted on a single chip \cite{natureHouck} and that disorder of resonator frequencies can be kept at a tolerable level for tens of resonators
\cite{arxivHouck}. Moreover, due to the size of the resonators, the Josephson junctions are spatially well separated and can be wired up individually which enables tuning each junction independently.
In addition to the typical many body features, the resonator network is predestined for studying the effects of decoherence and driving \cite{Hartmann2010,Carusotto,Schmidt,Marcos,Tomadin} as well as the transition from a single entity to a many body physics. In this sense, the system we propose can also be viewed as a quantum simulator. Notably, a set of experimental analysis methods for the quantum statistics of resonator networks have been demonstrated in recent years \cite{Menzel,LangEichler,Devoret}

The article is organized as follows. First we present two versions of one-dimensional arrays of coupled nonlinear transmission line resonators that can be described by a Bose-Hubbard Hamiltonian. In the next step we derive the full Hamiltonian for a single transmission line resonator intersected by a Josephson junction. We then focus our attention on the fundamental mode of the resonator, derive an approximate Hamiltonian for it that includes the junction nonlinearity and show that it represents a single site of a Bose-Hubbard Hamiltonian. Finally, regarding an experiment we quantify how much nonlinearity can be achieved with parameters of recently realized circuit QED setups.

\section{Chains of transmission line resonators}

\begin{figure}
\includegraphics{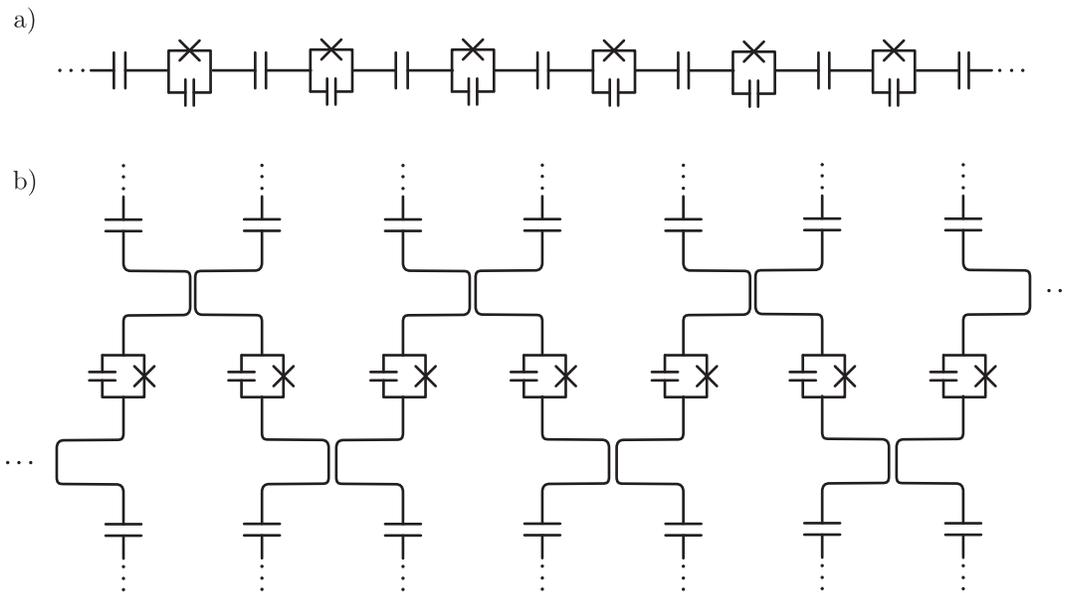}
\caption{Two different coupling schemes for one-dimensional networks of nonlinear superconducting transmission line resonators. a) Capacitive coupling of the transmission line resonators. b) Capacitive and inductive coupling with the advantage of individual addressability for the resonators. Only the central conductors of the transmission lines are drawn. Ground planes are omitted.}
\label{fig:ResChain}
\end{figure}
Circuit QED offers ample possibilities to couple several resonators to a network as there are basically no limitations on the topology and geometry of networks of transmission line resonators \cite{1367-2630-13-9-095008,PhysRevA.82.043811}, except for constraints imposed by the detection circuitry and by space on the chip. In \fref{fig:ResChain} we display two specific configurations of one-dimensional networks (chains) of transmission line resonators.
In \fref{fig:ResChain} a), the resonators are coupled at both ends \cite{leib}, which form a small capacitance and thereby enable photon hopping between adjacent resonators. In this way, also two-dimensional lattices can be formed by coupling more than two resonators at their ends \cite{PhysRevA.82.043811}. The advantage of this coupling scheme is scalability whereas a drawback results from the fact that one can only probe the output fields at the borders of the entire network \cite{leib}. However, transmission line resonators can also be coupled by reducing the distance between their central conductors for a certain length as shown in \fref{fig:ResChain} b). Depending on the exact position of the convergence, and the specific mode under consideration, the coupling between the two resonators is in general both capacitive and inductive \cite{mariantoni,reuther}. The advantage of this coupling scheme is that each transmission line resonator can be probed individually. This advantage however only holds for one-dimensional networks. 

For both designs the coupling constants are typically smaller than the frequencies of the field modes and therefore a rotating wave approximation is applicable. The coupling Hamiltonian consequently reads \cite{1367-2630-13-9-095008},
\begin{equation*}
\sum\limits_{\left\langle i,j\right\rangle}J_{i,j}\left(a_i a_j^{\dag}+a_i^{\dag}a_j\right)\,,
\end{equation*}
where the $a_i$ are the mode operators of the field modes in the respective transmission line resonator, the sum comprises all coupled transmission line resonators and $J_{i,j}$ incorporates all microscopic details of the interaction like the respective mode, the coupling capacitance and the coupling inductance. Based on present day precision in sample production we assume, for the following, all coupling constants to be equal, $J_{i,j}=J$. As shown below every nonlinear transmission line resonator can be modeled by a harmonic oscillator Hamiltonian with a Kerr nonlinearity $U$. For the whole chain of nonlinear transmission line resonators we therefore get a Bose-Hubbard Hamiltonian of the form,
\begin{equation}\label{eq:BHHamiltonian}
H_{BH}=\sum\limits_i\left[ \omega a_i^{\dag}a_i+ U a_i^{\dag}a_i^{\dag}a_ia_i+J\left(a_i^{\dag}a_{i+1}+a_ia_{i+1}^{\dag}\right)\right]
\end{equation}

\section{Nonlinear transmission line resonator}

In this section, we derive the mode spectrum of a single transmission line resonator that is intersected by a Josephson junction \cite{bourassa:032109}. To this end, we first disregard its nonlinearity and compute the normal modes of the harmonic part of its Hamiltonian.
The nonlinearity is then written in terms of these normal modes. To correctly identify the boundary conditions and the wave equation governing the flux function of the nonlinear resonator, we start by deriving the discretized, lumped-element representation of the transmission line resonator.

\begin{figure}
\includegraphics[keepaspectratio,width=5cm]{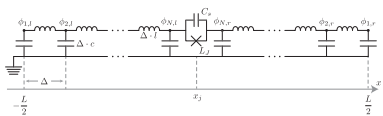}
\caption{Lumped-element circuit model for the nonlinear transmission line resonator. Every node in the circuit represents an element of finite length $\Delta$ of the transmission line which can be described by an inductance, $\Delta\cdot l$, and a capacitance with respect to the ground plane, $\Delta\cdot c$. Here, $l$ and $c$ are the inductance and capacitance per unit length of the transmission line. The Josephson junction, with inductance $L_J$, intersects the lined up inductances and provides a shunt capacitance $C_s$, which consists of the capacitance of the junction itself and a possible additional shunting capacitance.}
\label{fig:LumpedModel}
\end{figure}

\subsection{Lumped-element model}
We divide the resonator into finite length lumps that are represented in our circuit diagram by nodes. These nodes are inductively connected in a series. The inductance of each connection is $\Delta\cdot l$, the length of the unit cell $\Delta$ times the inductance per unit length $l$ of the transmission line. Each lump of transmission line is itself capacitively coupled to the ground plane. We represent this by a capacitance to ground at each node in the circuit diagram. The capacitance $\Delta \cdot c$ is again the product of the length of the unit cell $\Delta$ and the capacitance per unit length $c$. The Josephson junction interrupts the transmission line at the location $x_J$, where the two parts of the transmission line are connected capacitively  and through the overlap of their Cooper-pair wavefunctions. The latter is represented in the circuit diagram by a nonlinear Josephson inductance which is shunted by the capacitance of the Josephson junction itself and an additional shunting capacitance $C_s=C_J+C_{shun}$. The Lagrangian of the whole discretized transmission line setup reads,
\begin{equation*}
\mathcal{L} = \mathcal{L}_l + \mathcal{L}_r + \mathcal{L}_{JJ}
\end{equation*} 
where,
\begin{eqnarray*}
\mathcal{L}_{l/r}&=&\sum_{i=1}^N\frac{\Delta \cdot c}{2}\dot{\phi}_{i,l/r}^2-\sum_{i=1}^{N-1}\frac{\left(\phi_{i,l/r}-\phi_{i+1,l/r}\right)^2}{2\Delta\cdot l}\\
\mathcal{L}_{JJ}&=&\frac{C_s}{2}\left(\dot{\phi}_{N,l}-\dot{\phi}_{N,r}\right)^2+E_J\cos\left(\frac{2\pi}{\Phi_0}(\phi_{N,l}-\phi_{N,r})\right)\,,
\end{eqnarray*}
with $\Phi_0=h/2e$ the quantum of flux, $E_J$ the Josephson energy and $\phi_{i,l/r}=\int_{-\infty}^t V_{i,l/r}(t')dt'$ the node fluxes \cite{fluctuation}. $\mathcal{L}_{l/r}$ are the Lagrangians of the transmission lines to the left and the right of the Josephson junction respectively. These are represented by $N$ nodes each, with the first node being located at the respective end of the transmission line and the Nth node being the node next to the Josephson junction, see figure \ref{fig:LumpedModel}. $\mathcal{L}_{JJ}$ is the Lagrangian of the Josephson junction. Our goal is to find the eigenmodes for the linear part of the Hamiltonian which will be quantized  in the usual way.

We will proceed to accomplish this task as follows: We first separate the linear parts of the Langrangian from the rest, take the continuum limit and derive the wave equation for the linear part. Then we do a separation ansatz for the spatial and temporal dimension, identify the spatial boundary conditions and derive the corresponding eigenmodes. The time variation of the eigenmodes will then be described by harmonic oscillator equations of motion and therefore can be quantized in the same way as independent harmonic oscillators.
 
Since $\mathcal{L}_l$ and $\mathcal{L}_r$ are linear Langrangians, the separation of the linear and nonlinear parts of the Lagrangian can be accomplished by dividing the Lagrangian for the Josephson junction in the following way,
\begin{equation*}
\mathcal{L}_{JJ} = \mathcal{L}_{JJ}^{lin} + \mathcal{L}_{JJ}^{nonlin}
\end{equation*} 
where,
\begin{eqnarray*}
\mathcal{L}_{JJ}^{lin}&=&\frac{C_s}{2}\left(\dot{\phi}_{N,l}-\dot{\phi}_{N,r}\right)^2-\frac{1}{2L_J}\left(\phi_{N,l}-\phi_{N,r}\right)^2\\
 \mathcal{L}_{JJ}^{nonlin}&=&E_J\cos\left(\frac{2\pi}{\Phi_0}(\phi_{N,l}-\phi_{N,r})\right)+\frac{1}{2L_J}\left(\phi_{N,l}-\phi_{N,r}\right)^2\\
&=&E_J\left(1+\sum\limits_{n=2}^{\infty}\frac{-1^n}{(2n)!}\left(\frac{2\pi(\phi_{N,l}-\phi_{N,r})}{\Phi_0}\right)^{2n}\right)\,.
\end{eqnarray*}
Here $L_J=\Phi_0^2/(4\pi^2 E_J)$ is the Josephson inductance.
The linear part of the Lagrangian leads to the following set of equations of motion for the node fluxes. For the nodes at both ends of the transmission line the equations of motion read,
\begin{equation}\label{eq:discreteBorder}
\frac{-(\phi_{1,l/r}-\phi_{2,l/r})}{\Delta\cdot l} = \Delta \cdot c \ddot{ \phi}_{l/r,1}\,,
\end{equation}
the equations for the node fluxes at the Josephson junction are,
\begin{equation}\label{eq:discreteJoBorder}
\frac{(\phi_{N-1,l/r}-\phi_{N,l/r})}{\Delta\cdot l}\pm\frac{1}{L_J}(\phi_{N,l}-\phi_{N,r})=\Delta\cdot c \ddot{\phi}_{N,l/r}\pm C_s(\ddot{\phi}_{N,l}-\ddot{\phi}_{N,r})\,,
\end{equation}
and for every remaining node flux  we get,
\begin{equation}\label{eq:discreteWaveEq}
\frac{\phi_{i-1,l/r}-2\phi_{i,l/r}+\phi_{i+1,l/r}}{\Delta^2}v^2=\ddot{\phi}_{i,l/r}\,,
\end{equation}
where $v=1/\sqrt{lc}$ is the phase velocity in the transmission line.
\subsection{Continuum limit}
To take the continuum limit, we shrink the size, $\Delta$, of our unit-cell of transmission line  and transform every sum in the Lagrangian into an integral over the spatial coordinate, $\mathcal{L}=\sum(...)\Delta\to \mathcal{L}=\int(...)dx$. Accordingly, differences between neighboring node fluxes divided by the size of our unit cell become derivatives with respect to the spatial dimension. We note that the node fluxes to the right and the left of the Josephson junction can have a finite difference in the continuum limit which is a consequence of our model assumption that the Josephson junction has no spatial extension in  the direction of the transmission line. This is a valid assumption as long as we study spatial modes of a wavelength that is much larger than the size of the Josephson junction. In this way of downsizing the unit cell, the inductance of the Josephson junction remains finite, even though its extension becomes vanishingly small. The equations of motion for the fluxes at discontinuities of the transmission line \eref{eq:discreteJoBorder} and \eref{eq:discreteBorder} provide us with the boundary conditions,
\begin{eqnarray}
\left.-\partial_x\phi_{l/r}(x)\right |_{x=\mp\frac{L}{2}}&=&0\label{eq:BoundaryCondition1}\\
\left.-\partial_x\phi_l(x)\right |_{x=x_J}&=&-\left.\partial_x\phi_r(x)\right |_{x=x_J}=lC_s\delta\ddot{\phi}+\frac{l}{L_J}\delta\phi\label{eq:BoundaryCondition2}\,.
\end{eqnarray} 
The first boundary condition \eref{eq:BoundaryCondition1} is a consequence of charge conservation. The current, which is the spatial derivative of the flux, has to vanish at both ends of the transmission line. The second boundary condition \eref{eq:BoundaryCondition2} again states the fact of charge conservation, imposing that the current flowing into the Josephson junction at point $x_J$ on the right hand side has to exit the Josephson junction on the left hand side. This current that flows through the Josephson junction is in turn related to the phase drop $\delta\phi = (\left.\phi_l\right|_{x=x_J}-\left.\phi_r\right|_{x=x_J})$ across the junction. The equations of motion for the node fluxes in the remaining nodes of the transmission line resonator yield the wave equation,
\begin{equation}\label{eq:wave}
\partial_x^2\phi_{l/r}v^2=\partial_t^2\phi_{l/r}\,,
\end{equation}
for the flux $\phi_{l/r}$.

\subsection{Eigenmodes of the linearized transmission line resonator}

To find the eigenmodes of the linearized transmission line resonator we use an ansatz with separation of variables, 
\begin{equation}\label{eq:SepAnsatz}
\phi_{l/r}(x,t)=f_{l/r}(x)g(t)\,,
\end{equation}
and derive appropriate spatial normal modes that fulfill the wave equation \eref{eq:wave}, in combination with the boundary conditions \eref{eq:BoundaryCondition1} and \eref{eq:BoundaryCondition2}. The wave equation \eref{eq:wave} leads to the following equations for $f_{l/r}$ and $g$,
\begin{eqnarray*}
f_{l/r}^{''}(x)+k^2f_{l/r}(x)&=&0\\
\ddot{g}(t)+\omega^2 g(t)& =&0\,,
\end{eqnarray*}
where $\omega$ and $k$ are the frequency and wave vector of the normal mode which are connected by the linear dispersion relation $|k|v=\omega$. The boundary conditions at both ends of the transmission line resonator directly translate into boundary conditions for the function $f(x)$.
To ensure current conservation at both ends of the transmission line resonator \eref{eq:BoundaryCondition1} we choose the functions $f_l$ and $f_r$ to read,
\begin{eqnarray}
f_l(x)&=&A_l\cos(k(x+\frac{L}{2}))\label{eq:SpatialModeAnsatz1}\\
f_r(x)&=&A_r\cos(k(x-\frac{L}{2}))\,.\label{eq:SpatialModeAnsatz2}
\end{eqnarray}
We now restrict our calculations to the case where the Josephson junction is in the middle of the transmission line resonator, $x_J=0$, and hence current conservation through the Josephson junction, compare equation \eref{eq:BoundaryCondition2}, requires,
\begin{equation*}
\left.\partial_x f_l(x)\right|_{x=0}=-A_lk\sin(k\frac{L}{2})=A_rk\sin(k\frac{L}{2})=\left.\partial f_r(x)\right|_{x=0}\,.
\end{equation*}
There are two different ways to fulfill this boundary condition. Choosing $k_n= 2\pi n /L$ provides us with symmetric (even) spatial modes that do not couple to the Josephson junction because no current is flowing through it. Alternatively we can choose $A_l=-A_r$ which provides us with the antisymmetric (odd) spatial modes. We specifically choose $A_l=-A_r=1$ and consequently obtain orthogonal but not normalized modes. The remaining work to be done is to connect the current flowing through the Josephson junction and the phase drop across the Josephson junction with the help of boundary condition \eref{eq:BoundaryCondition2}, where we insert our separation ansatz \eref{eq:SepAnsatz} and get,
\begin{equation*}
-f_l^{'}(0)g(t)=-lC_s\omega^2 (f_l(0)-f_r(0))g(t)+\frac{l}{L_J}(f_l(0)-f_r(0))g(t)\,.
\end{equation*}
Here, we used the equation of motion for the function $g(t)$. With our ansatz for the spatial modes, equations \eref{eq:SpatialModeAnsatz1} and \eref{eq:SpatialModeAnsatz2}, we get the transcendental equation for the spatial mode frequencies $\omega_n$,
\begin{equation}
\frac{\omega_n}{v}=\cot\left(\frac{\omega_n}{v}\frac{L}{2}\right)\,\frac{2l}{L_J}\,\left(1-\frac{\omega_n^2}{\omega_p^2}\right)\,,\label{eq:EV}
\end{equation}
where $\omega_p = 1/\sqrt{L_JC_s}$ is the plasma frequency of the capacitively shunted Josephson junction.
With the help of this transcendental equation we compute the frequencies for the odd modes of the system with $n=2k+1$. For the even modes the current through the Josephson junction vanishes and consequently the flux drop across the Josephson junction also vanishes.

One can check the orthogonality of the odd modes by integrating the product of two different ($m\neq n$) odd modes over the length of the transmission line resonator, 
\begin{equation}\label{eq:normalize1}
\int_{-\frac{L}{2}}^{\frac{L}{2}}f_n(x)f_m(x)=\frac{2\left(k_m\cos(k_n\frac{L}{2})\sin(k_m\frac{L}{2})-k_n\cos(k_m\frac{L}{2})\sin(k_n\frac{L}{2})\right)}{k_m^2-k_n^2}\,.
\end{equation}
Using the transcendental equation for the eigenfrequencies \eref{eq:EV} we get,
\begin{equation*}
\frac{2\left(k_m\cos(k_n\frac{L}{2})\sin(k_m\frac{L}{2})-k_n\cos(k_m\frac{L}{2})\sin(k_n\frac{L}{2})\right)}{k_m^2-k_n^2}=-C_s\delta f_n\delta f_m\,,
\end{equation*}
where $\delta f_n$ is the flux drop of the spatial eigenmode $n$, $\delta f_n = f_{n,l}(0)-f_{n,r}(0)$.
This result leads us to alter the usual $l_2$ scalar product to account for the discontinuity at the Josephson junction,
\begin{eqnarray}\label{eq:orthoRel}
c\int_{-\frac{L}{2}}^{\frac{L}{2}}f_n(x)f_m(x)dx+C_s\delta f_n \delta f_m &=& \delta_{n,m} \eta_n\\
\eta_n=c\left(\frac{L}{2}+\frac{\delta f_n^2}{k_n^2}\frac{l}{2L_J}\left(1+\frac{\omega_n^2}{\omega_p^2}\right)\right)\,.\nonumber
\end{eqnarray}
Here, we opted not to normalize the spatial eigenmodes but rather regard them as modes with different effective masses $\eta_n$.  
Now we can reformulate the flux field in terms of the eigenmodes $f_n$,
\begin{equation*}
\phi(x,t)= \sum_{n=1}^{\infty}g_n(t)f_n(x)\,,
\end{equation*}
and with the help of the orthogonality relation \eref{eq:orthoRel} we rearrange the integral in the Lagrange function into a sum over the infinite number of spatial eigenmodes,
\begin{equation*}
\mathcal{L}=\sum_{n=1}^{\infty}\left(\frac{\eta_n}{2}\dot{g}_n^2-\frac{1}{2}\eta_n\omega_n^2 g_n^2\right)+\left(E_J\cos\left(\frac{2\pi \delta\phi}{\Phi_0}\right)+\frac{1}{2L_J}\delta \phi^2\right)\,,
\end{equation*}
where $\delta\phi=\sum_n g_n(t)\delta f_n$. Here, we already neglected the even spatial modes, because they are completely decoupled from the odd modes. The corresponding Hamilton function can be found by a Legendre transformation which leaves the nonlinear part of the Lagrangian unaltered,
\begin{eqnarray*}
H&=&\sum_{n=1}^{\infty}\left(\frac{\pi_n^2}{2\eta_n}+\frac{1}{2}\eta_n \omega_n^2g_n^2\right)-\left(E_J\cos\left(\frac{2\pi\delta\phi}{\Phi_0}\right)+\frac{1}{2L_J}\delta\phi^2\right)\\
\pi_n&=&\eta_n\dot{\phi}_n\,. 
\end{eqnarray*}
We quantize the theory in the usual way by imposing canonical commutation relations $\left[\hat{\pi}_n,\hat{g}_m\right]=-i\delta_{n,m}$. The generalized coordinates $\hat{g}_n$ and momenta $\hat{\pi}_n$ can be expressed in terms of annihilation and creation operators $a_n$ and $a_n^{\dag}$ via the relations,
\begin{eqnarray*}
\hat{\pi}_n&=&-i\sqrt{\frac{\eta_n\omega_n}{2}}(a_n-a_n^{\dag})\\
\hat{g}_n&=&\frac{1}{\sqrt{2\eta_n\omega_n}}(a_n+a_n^{\dag})\,,
\end{eqnarray*}
with $\left[a_n,a_m^{\dag}\right]=\delta_{n,m}$.
Finally, we arrive at the following Hamilton operator,
\begin{eqnarray}
H&=&\sum_{n=1}^{\infty}\omega_n(a_n^{\dag}a_n+\frac{1}{2})+H^{\text{nonlin}}\nonumber\\
H^{\text{nonlin}}&=&-E_J\left(\cos\left(\delta\tilde{\phi}\right)+\frac{1}{2}\delta\tilde{\phi}^2\right)\,,\label{eq:unHarmHam}
\end{eqnarray}
with 
\begin{equation*}
\delta\tilde{\phi}=\sum_{n=1}^{\infty}\lambda_n\left(a_n+a_n^{\dag}\right)\quad \text{and,} \quad
\lambda_n=\frac{2\pi\delta f_n}{\Phi_0\sqrt{2\eta_n \omega_n}}\,.
\end{equation*}

\subsection{Derivation of the Kerr-nonlinearity}
In principle, every odd mode is coupled to all other odd modes by the nonlinear Hamiltonian \eref{eq:unHarmHam}. 
As we illustrate in the following, the interactions between different modes are however negligible in the low energy limit. Firstly, we note that the nonlinear part of the Hamiltonian \eref{eq:unHarmHam} does not contain any hopping terms of the form $a_ia_j^{\dag}+a_i^{\dag}a_j$ for $i\not=j$ because we already diagonalized the complete linear part of the Hamiltonian. Secondly, density-density interactions of the type $a_i^{\dag}a_i a_j^{\dag}a_j$ can be neglected because we assume all the modes but the fundamental mode to be in the vacuum state and the negligible interaction between the modes prevents them from becoming populated. Finally, the remaining higher order terms do not contribute a significant coupling. To separate the Hamilton function of the fundamental mode from the rest, we apply a rotating wave approximation, the validity of which we confirm below. This approximation allows us to substantially simplify the cosine term of the nonlinear part of the Hamilton operator \eref{eq:unHarmHam}. To this end we, consider the identity,
\begin{eqnarray*}
\cos\left(\sum\limits_{n=1}^{\infty} \lambda_n(a_n+a_n^{\dag})\right)&=&\cos\left(\lambda_1(a_1+a_1^{\dag})\right)\cos\left(\sum\limits_{n=2}^{\infty} \lambda_n(a_n+a_n^{\dag})\right)\\
&&-\sin\left(\lambda_1(a_1+a_1^{\dag})\right)\sin\left(\sum\limits_{n=2}^{\infty} \lambda_n(a_n+a_n^{\dag})\right)\,.
\end{eqnarray*} 
The sine terms will only contain operator products with odd powers of the individual mode operators, hence the second addend can be neglected as part of the rotating wave approximation. If we iterate this consideration, we can approximate the cosine of the sum of our mode position operators by the product of the cosines of the mode position operators,
\begin{equation*}
\cos\left(\sum_n \lambda_n (a_n+a_n^{\dag})\right)\to \prod_n \cos\left(\lambda_n(a_n+a_n^{\dag})\right)\,.
\end{equation*}
The cosine terms can further be approximated by a sum of normally ordered operator products with equal numbers of annihilation and creation operators,
\begin{equation}\label{eq:CosineExp}
\cos\left(\lambda_n (a_n+a_n^{\dag})\right)\to  \alpha_{n,0}+\alpha_{n,1} a_n^{\dag} a_n+\alpha_{n,2} a_n^{\dag}a_n^{\dag} a_n a_n+\dots\,.
\end{equation}
The expansion coefficients $\alpha_{n,k}$ in equation \eref{eq:CosineExp} are not the usual expansion coefficients of the cosine since one has to take into account that also higher powers of $\lambda_n(a_n+a_n^{\dag})$ generate operator products of lower powers if we attempt to normal order the mode operator binomials. With the formula for mode operator binomials given in \cite{0305-4470-16-16-019},
\begin{equation*}
(a+a^{\dag})^m=\sum\limits_{k=0}^{[m/2]}\sum\limits_{i=0}^{m-2k}\frac{m!(a^{\dag})^i a^{m-2k-i}}{2^k k! i! (m-2k-i)!}\,,
\end{equation*}
where $[m/2]$ denotes the largest integer less than or equal to $m/2$, we can identify the prefactors for the rotating wave expansion of the cosine,
\begin{equation*}
\cos\left(\lambda_n(a_n+a_n^{\dag})\right)\to e^{-\frac{\lambda_n^2}{2}}\left(1-\lambda_n^2 a_n^{\dag}a_n+\frac{\lambda_n^4}{4} a_n^{\dag} a_n^{\dag}a_na_n+\dots\right)\,.
\end{equation*}
We assume that only the fundamental mode is populated and can therefore neglect all modes with $n\not=1$ in $H^{\text{nonlin}}$. Since $|\lambda_{n}| \ll 1$, as will be confirmed below, and because we are only interested in low photon numbers where $\lambda_{n}^{2} \langle a_n^{\dag}a_n \rangle \ll 1$, we can truncate the cosine expansion after the quartic order term. We thus get for the nonlinear part of our Hamilton operator in the rotating wave and low photon number approximation,
\begin{equation*}
H^{\text{nonlin}}\to-E_J\left(\left(1-\lambda_1^2 a_1^{\dag}a_1+\frac{\lambda_1^4}{4} a_1^{\dag} a_1^{\dag}a_1a_1\right)\prod_{n=1}^{\infty}e^{-\frac{\lambda_n^2}{2}}+\lambda_1^2 a_1^{\dag}a_1\right)\,.
\end{equation*} 
Any constant in the Hamilton operator provides us with an unmeasurable shift of the overall energy, therefore we may omit all constant terms and get the following Hamiltonian for the fundamental mode in rotating wave approximation,
\begin{equation*}
H_1=\left(\omega_1-\delta \omega\right) a_1^{\dag}a_1-U_{1} a_1^{\dag}a_1^{\dag}a_1a_1\,,
\end{equation*}
with
\begin{eqnarray*}
\delta \omega_{1} &=& E_J \lambda_1^2\left(1-\prod\limits_{n=1}^{\infty} e^{-\frac{\lambda_n^2}{2}}\right)\\
U_{1} &=& E_J \frac{\lambda_1^4}{4}\prod\limits_{n=1}^{\infty} e^{-\frac{\lambda_n^2}{2}}\,,
\end{eqnarray*}
where $\delta\omega_{1}$ is a small renormalization  of the fundamental mode due to the nonlinearity.

\section{Experimental prerequisites}

In this section we discuss how one would realize the above introduced nonlinear transmission line resonator. We propose to use a dc SQUID instead of the single Josephson junction in order to have a tunable and large nonlinearity $U_{1}$. If the self inductance of the SQUID is negligible, it can be modeled as a single Josephson junction whose critical current $I_c=2I_0|\cos(\Phi/(2\Phi_0))|$ (cf. \cite{clarkesquid}) can be tuned via the external flux $\Phi$ threading the SQUID loop. 

The further discussion will be twofold. We first consider the experimental limitations which are imposed on our theoretical model by the physics of the superconducting circuitry and then discuss how the setup should be tuned to realize a sufficient amount of nonlinearity.
In an experiment, it is favorable that the fundamental mode of the nonlinear resonator is within a frequency band suitable for state-of-the-art microwave measuring devices. Furthermore, the thermal population should be insignificant at least at the millikelvin temperatures used in most experiments. We therefore choose the length $L$ of the transmission line resonator such that for all relevant critical currents $I_c$ the frequency of the fundamental mode is in the range $\unit{4}{\giga\hertz}<\omega/(2\pi) <\unit{8}{\giga\hertz}$. All other design parameters of the transmission line ($Z_0$ , $l$, $c$) are typical values found in many circuit QED setups. The critical current $I_c$ of the SQUID can be varied in two different ways. Either by design, or, as sketched above, by applying a flux bias to the SQUID loop. In this work we study a wide range of critical currents, particularly focussing on the experimentally well accessible  range of $\unit{8 \cdot 10^{-7}}{\ampere}<I_c<\unit{4\cdot 10^{-6}}{\ampere}$. A summary of all relevant parameters is given in \tref{tab:ExpParam}. 

\begin{table}
\caption{\label{tab:ExpParam} }
\begin{indented}
\item[] \begin{tabular}{@{}rll}
\br
\multicolumn{3}{l}{transmission line} \\
\mr
\hspace{10mm} characteristic impedance & $ Z_0 $ & $\unit{50}{\ohm}$\\
phase velocity & $v$ & $\unit{0.94 \cdot 10^8}{\meter\per\second}$\\
inductance per length & $l$ & $\unit{5\cdot 10^{-7}}{\henry\per\meter}$\\
capacitance per length & $c$ & $\unit{2 \cdot 10^{-10}}{\farad\per\meter}$\\
\br
\multicolumn{3}{l}{dc SQUID}\\
\mr
shunting capacitance&$C_s$& $\unit{2 \cdot 10^{-12}}{\farad}$\\
maximal critical current range&$I_c$&$\unit{1\cdot 10^{-7}}{\ampere}-\unit{1 \cdot 10^{-4}}{\ampere}$\\
\br
\end{tabular}
\end{indented}
\end{table}

In order to  solve the transcendental equation for the spatial mode frequencies \eref{eq:EV} we first reformulate it in terms of the phase $\varphi$ associated to the wave vector of the spatial eigenmode, $\varphi=k (L /2)$, 
\begin{equation}\label{eq:EV2}
\frac{\varphi}{1-\frac{\varphi^2}{\varphi_p^2}}\frac{L_J}{l L}=\cot(\varphi)\,,
\end{equation}
where $\varphi_p$ is the corresponding phase for the capacitively shunted dc-SQUID, $\varphi_p = (\omega_p/v) (L/2)$. Every time the left hand side of this equation equals the right hand side for some specific $\varphi=\varphi_n$, we have found one valid mode frequency $\omega_n=\varphi_n 2v/L$. For our choice of parameters, the right hand side has a narrow divergence for $\varphi = \varphi_p$ respectively $\omega = \omega_p$ and quickly approaches zero for values of $\varphi$ unequal to $\varphi_p$. Consequently, we have two different possibilities for approximate solutions of the transcendental equation. For  $\varphi\neq\varphi_p$, we can approximate the left hand side of the transcendental equation by $0$ and find,
\begin{eqnarray*}
\varphi_n &=& (2n+1)\frac{\pi}{2} \quad n \in \mathbb{N}\\
\omega_n &=& (2n +1)\frac{\pi v}{L}\quad n \in \mathbb{N}\,.
\end{eqnarray*}
This corresponds to the situation where the plasma frequency of the capacitively shunted dc-SQUID $\omega_p$ is off-resonant with the unperturbed frequency of the transmission line resonator mode and consequently we recover the unperturbed frequencies of the resonator without dc-SQUID. The situation changes if one searches for system modes near the plasma frequency of the capacitively shunted dc-SQUID, $\varphi\approx \varphi_p$. Then, the divergence of the left hand side becomes infinitely narrow  and the solution of the transcendental equation is approximately $\varphi_n\approx \varphi_p$. These considerations are confirmed by the numerically calculated mode spectrum of the linear part of the combined device. Here, we find the spectrum of the odd modes of the unperturbed transmission line resonator modes and the plasma frequency of the capacitively shunted dc-SQUID with avoided crossings near $(2n +1)\frac{\pi v}{L}\approx\omega_p$. The latter arise from their mutual interaction.

\begin{figure}
\includegraphics{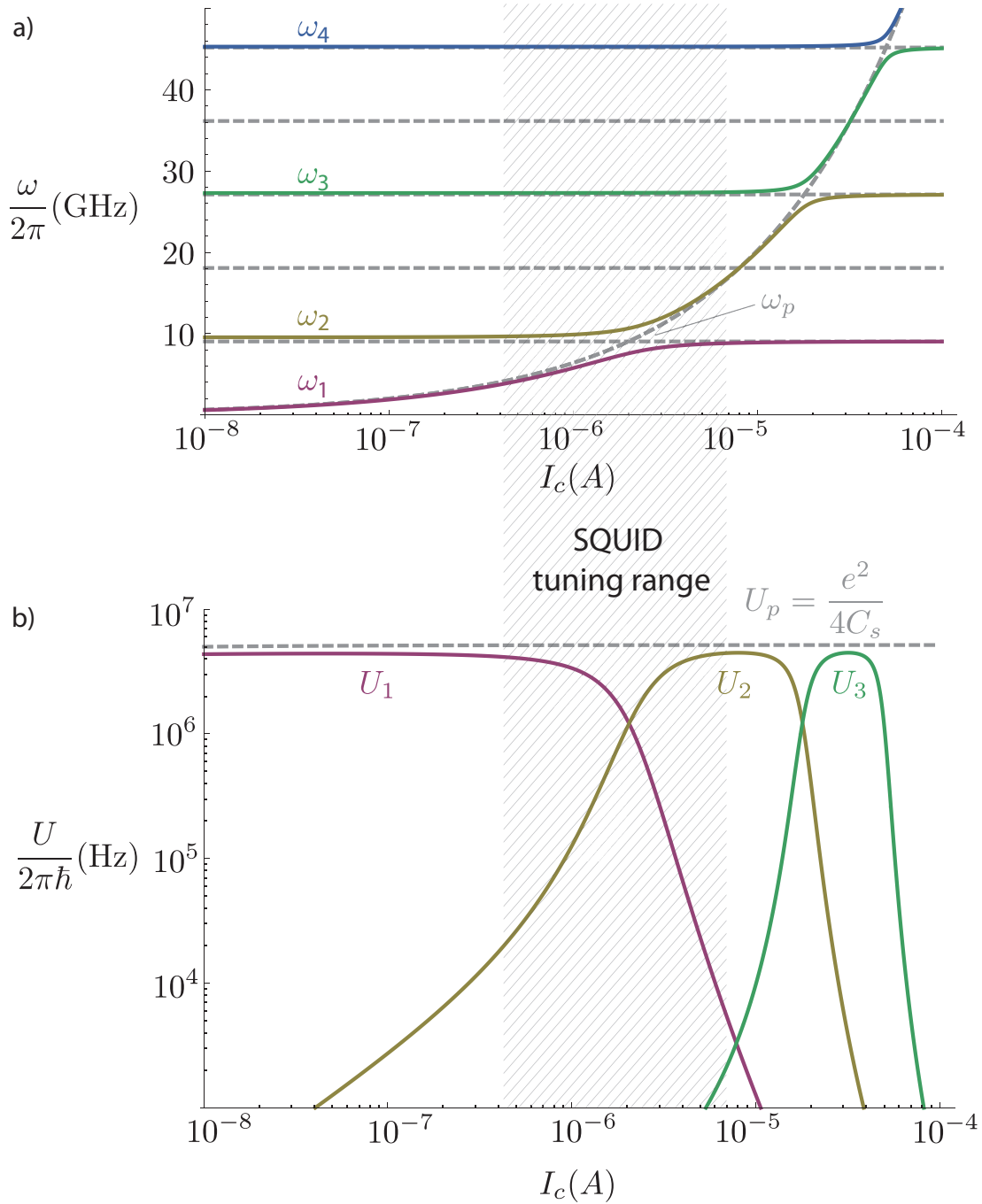}
\caption{a) Numerically calculated frequencies of the linear part of the transmission line resonator intersected by the capacitively shunted dc-SQUID. Anti-crossings show up where the plasma frequency of the capacitively shunted dc-SQUID $\omega_p$ (dashed ascending line) matches one of the frequencies of the odd modes of the transmission line resonator (dashed horizontal lines). b) Nonlinearity parameter of the different modes $U_n$. If the frequency of the mode is near the plasma frequency of the capacitively shunted dc-SQUID its nonlinearity $e^2/(4 C_s)$ is inherited by the mode.}
\label{fig:specNonlin}
\end{figure}
To calculate the nonlinearity parameter for the fundamental mode, $U_1$, we need to evaluate the formally infinite product, $\prod\limits_{n=1}^{\infty}e^{-\frac{\lambda_n^2}{2}}$. Here the superconducting gap provides a natural cut-off frequency. If the first excited state of a mode of the transmission line resonator exceeds the superconducting gap in energy, Cooper pairs will break and strong dissipation processes will start. Therefore we only include modes that fit energetically into the superconducting gap in the above product and get $\prod\limits_{n=1}^{n_{\text{cutoff}}}e^{-\frac{\lambda_n^2}{2}}$. Here, $n_{\text{cutoff}}$ is the largest of all $n$ that fulfills the equality,
\begin{equation*}
\hbar\frac{\pi v}{L}(2n-1)<\Delta(T)\quad\text{and}\quad\Delta(T)=3.52 \, k_B T_c \sqrt{1-\frac{T}{T_c}} \,,
\end{equation*}
with $\Delta$ the energy gap of the superconductor. Here we used the critical temperature of niobium. Niobium technique is favorable because of its high transition temperature of $9.2$ Kelvin that enables fast precharacterization at liquid helium temperatures before cooling the whole device down to the millikelvin range.

The nonlinearity of the fundamental mode is inherited from the capacitively shunted dc-SQUID for values of the critical current $I_c$ where the frequency of the fundamental mode is near the plasma frequency. This principle of inheritance, where a collective mode of the composite resonator-junction system adopts the properties of one of its constituents if its frequency matches the frequency of the respective isolated constituent, is further illustrated by the nonlinearity $U_n$ of higher modes (see \fref{fig:specNonlin}). The nonlinearities for these modes are calculated in the same manner as the nonlinearity for the fundamental mode, $U_1$, where we performed a rotating wave approximation and assumed all other modes to be in the vacuum state.
Here we can see that also higher modes could be employed as onsite nonlinear mode as they acquire the same amount of nonlinearity.
In \fref{fig:specNonlin}, we hatched a proposed tuning range for the SQUID critical current of $\unit{8 \cdot 10^{-7}}{\ampere}<I_c<\unit{4\cdot 10^{-6}}{\ampere}$. This results in a tuning range for the nonlinearity $U_1$ of $U_{min}/(2\pi \hbar)=\unit{10}{\kilo\hertz}$ to $U_{max}/(2\pi \hbar)=\unit{4}{\mega\hertz}$ and a frequency drag for the fundamental mode of $\omega_{min}/(2\pi)=\unit{5.2}{\giga\hertz}$ to $\omega_{max}/(2\pi)=\unit{8.5}{\giga\hertz}$. This provides us with sufficient nonlinearity to exceed present day decay rates of transmission line resonators $\gamma/(2\pi)\approx \unit{100}{\kilo\hertz}$ while still keeping the frequency drag within the bandwidth of state of the art microwave detection.

These values allow one to investigate the Bose-Hubbard Hamiltonian of \eref{eq:BHHamiltonian} with in situ tunability between the onsite nonlinearity dominated regime $U>J$ and the hopping dominated regime $J>U$ using present day technology. Finally, we note that the nonlinearity $U_{max}$ is here bound from above by the charging energy (see \fref{fig:specNonlin}). Therefore, it could be further increased by decreasing the shunting capacitance $C_s$. However, this would shift the anticrossing marking the region of tunability, and hence the operating interval, to lower critical currents, finally resulting in fabrication and noise issues. 

In the whole presented range of critical currents $\unit{8 \cdot 10^{-8}}{\ampere}<I_c<\unit{4\cdot 10^{-4}}{\ampere}$ we checked the validity of our rotating wave approximation. The most dominant process of all neglected ones is the exchange of three fundamental mode excitations with one excitation of the next higher odd mode, $a_1 a_1 a_1 a_2^{\dag}$. The prefactor for this term in the Hamiltonian is of the order, $E_J\lambda_1^3\lambda_2$. This should be compared to the frequency difference between three times the fundamental mode frequency $\omega_1$ and the second odd mode frequency $\omega_2$. For the whole considered range of critical currents the ratio of the prefactor and the frequency difference is vanishingly small, which validates our assumption of decoupled field modes.

\section{Summary}
In summary, we have introduced a scheme to study interacting quantum many-body systems formed by photons. Our approach employs a network of transmission line resonators which are themselves nonlinear. Hence, no qubit circuits are required to generate a nonlinearity for microwave photons. Our approach thus reduces the amount of control circuitry required for building the cavity network.
This reduced complexity is expected to enhance the perspectives for scaling up the technology to large networks of coupled circuit cavities.  
In combination with significant nonlinearities tunable between $\unit{10}{\kilo\hertz}$ and $\unit{4}{\mega\hertz}$, our system is a promising candidate for building quantum simulators for interesting few- and many-body physics problems.

\section*{Acknowledgements}
We acknowledge valuable discussions with J.Bourassa and K.F. Wulschner. This work is part of the collaborative research centre SFB 631 and the Emmy Noether project HA 5593/1-1, both funded by the German Research Foundation (DFG). F.D., A.M., and R.G. further acknowledge support from the German excellence initiative "Nanosystems Initiative Munich" (NIM), the EU Marie Curie initial training network CCQED, and the EU Specific Targeted Research Program PROMISCE.

\noappendix

\end{document}